\documentclass[aps,prb,reprint,twocolumn,superscriptaddress,showpacs]{revtex4-1}

\usepackage{graphicx}
\usepackage{amssymb}
\usepackage[colorlinks=true,linkcolor=blue,citecolor=blue,urlcolor=blue,filecolor=blue]{hyperref}

\begin{document}

\title{Determination of the bandgap and split-off band of wurtzite GaAs}

\author{Bernt Ketterer} 
\affiliation{Laboratoire des Mat\'{e}riaux Semiconducteurs, Institut des Mat\'{e}riaux, Ecole Polytechnique F\'{e}d\'{e}rale de Lausanne, CH-1015 Lausanne, Switzerland}

\author{Martin Heiss} 
\affiliation{Laboratoire des Mat\'{e}riaux Semiconducteurs, Institut des Mat\'{e}riaux, Ecole Polytechnique F\'{e}d\'{e}rale de Lausanne, CH-1015 Lausanne, Switzerland}

\author{Marie J. Livrozet} 
\affiliation{Laboratoire des Mat\'{e}riaux Semiconducteurs, Institut des Mat\'{e}riaux, Ecole Polytechnique F\'{e}d\'{e}rale de Lausanne, CH-1015 Lausanne, Switzerland}

\author{Elisabeth Reiger} 
\affiliation{Institute for Experimental and Applied Physics, University of Regensburg, Universit\"{a}tsstrasse 31, D-93053 Regensburg, Germany}

\author{Anna Fontcuberta i Morral}
\email{anna.fontcuberta-morral@epfl.ch}

\affiliation{Laboratoire des Mat\'{e}riaux Semiconducteurs, Institut des Mat\'{e}riaux, Ecole Polytechnique F\'{e}d\'{e}rale de Lausanne, CH-1015 Lausanne, Switzerland}

\date{\today}

\begin{abstract} 
GaAs nanowires with a 100\% wurtzite structure are synthesized by the vapor-liquid-solid method in a molecular beam epitaxy system, using gold as a catalyst. We use resonant Raman spectroscopy and photoluminescence to determine the position of the crystal-field split-off band of hexagonal wurtzite GaAs.  The temperature dependence of this transition enables us to extract the value at 0\,K, which is 1.982\,eV. Our photoluminescence excitation spectroscopy measurements are consistent with a band gap of GaAs wurtzite below 1.523\,eV.
\end{abstract}

\maketitle

Nanowires are filamentary crystals with a diameter of the order of few nanometers. Their increasing importance in both science and engineering is a consequence of the great number of novel experiments and applications they enable.\cite{ref1,ref2,ref3,ref4,ref5,ref6,ref7} It has been predicted and shown that the reduced diameter of nanowires allows the combination of lattice mismatched materials when they are fabricated in the nanowire form.\cite{ref8,ref9} The possibility of obtaining new material combinations opens great perspectives for example in the area of multiple junction photovoltaics.\cite{ref10} Recently, a new degree of freedom in the formation of heterostructures has appeared. The new type of heterostructure concerns the variation of the crystal phase along the nanowire instead of the material composition.\cite{ref11,ref12,ref13} The degree of control over the crystal phase can be astonishingly accurate depending on the growth method,\cite{ref14,ref15,ref16} so that perspectives for new device concepts are exciting the nano-science and nanotechnology community.

While the structural control is becoming increasingly sophisticated, few experimental reports have focused on the details of the electronic structure of wurtzite arsenides or phosphides. Recently, two groups applied photoluminescence excitation to determine the valence band structure of wurtzite InP.\cite{ref17,ref18} The results agreed well with the theoretical expectations. Wurtzite GaAs has shown to be more controversial. First, there are significant disagreements between the theoretical calculations of the bandgap.\cite{ref19,ref20}  Second, luminescence studies of different groups have shown results consistent with a bandgap of 1.54,\cite{ref21} 1.522,\cite{ref22} and 1.50\,eV.\cite{ref23,ref24,ref25} The apparent lack of agreement between the various groups might be explained by the fact that the optical and structural characterizations were not performed on exactly the same nanowire. Recently, we designed an experiment in which both the luminescence and electron microscopy measurements were realized on the identical nanowire.\cite{ref25} We investigated both nanowires presenting a mixture of wurtzite and zinc-blende and a 100\% of wurtzite crystal phase. These and previous experiments were consistent with a bandgap of 1.50\,eV for wurtzite GaAs.\cite{ref24,ref25}  

\begin{figure} \includegraphics[width=\columnwidth]{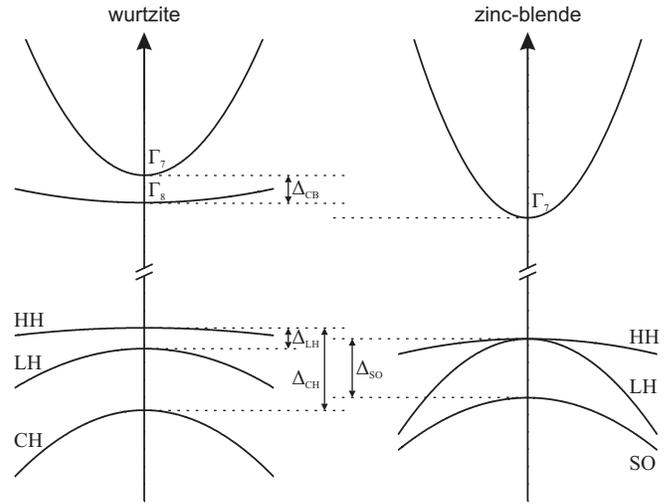}
\caption{\label{figure1}(left) Schematic band diagram for wurtzite GaAs near the Brillouin zone center according to Ref.~\onlinecite{ref19}. (right) Schematic band diagram for zinc-blende GaAs near the Brillouin zone center.	
} \end{figure}

As a consequence of the hexagonal symmetry, it has been shown that the band structure of wurtzite semiconductors exhibits important differences compared to the band structure of the respective zinc-blende (cubic) counterparts.\cite{ref19} In Fig.\,\ref{figure1} we compare the band structure of zinc-blende and wurtzite GaAs close to the $\Gamma$ point according to recent theoretical results from De and Pryor.\cite{ref19} As a consequence of the zone folding induced by the doubling of the unit cell along the (111) direction an additional conduction band with $\Gamma_8$ symmetry appears for the wurtzite structure. In contrast to other III-V semiconductors the energy separation $\Delta_{CB}$ between these two conduction bands is expected to be the smallest for the case of wurtzite GaAs. The theoretical predicted values of $\Delta_{CB}=-23$\,meV,\cite{ref20} $\Delta_{CB}=+85$\,meV,\cite{ref19} or $\Delta_{CB}=+87$\,meV\cite{ref25} are even smaller than the predicted splitting of the two uppermost valence bands. There, the crystal field splitting and spin-orbit interaction lift the degeneracy of the heavy and light hole states for the wurtzite structure.\cite{ref19,ref26} Furthermore a crystal-split-off hole (CH) band is predicted further down in energy below the valence band edge compared to the split-off band in zinc-blende GaAs.\cite{ref19} To the best of our knowledge, there are to date no studies providing the values of either the crystal splitting or split-off band for the case of wurtzite GaAs.

Luminescence studies allow the probing of transitions between the conduction band minimum and the highest energy valence band states. In order to obtain information on the valence band structure, i.e. crystal and split-off band splitting, other type of experiments such as photoluminescence excitation and resonant Raman scattering should be implemented.\cite{ref17,ref27,ref28,ref29} In this paper we use resonant Raman scattering and photoluminescence excitation spectroscopy to probe the crystal-field (CH) split-off valence band to conduction band transition and to provide more clarity and consistence in recent luminescence studies that attributed the bandgap of wurtzite to be at 1.50\,eV.

Wurtzite GaAs nanowires were grown by the Au-catalyzed Vapor-liquid-solid method on GaAs ($1\bar{1}1$)B substrates at a growth temperature of 540\,$^{\circ}$C under a As$_4$ Beam flux of $1.27 \cdot 10^{-6}$\,Torr at a Ga rate equivalent to a planar growth of 0.4 \AA/s. The growth time was 4 hours. The nucleation and growth followed the Vapor-Liquid-Solid mechanism, with Au as catalyst.\cite{ref30} Details on the growth procedure are described in.\cite{ref31} After the axial growth of the nanowires the growth parameters were changed to conditions suitable for planar growth and the nanowires were passivated by an epitaxial prismatic shell of AlGaAs/GaAs material.\cite{ref32} The 2D equivalent amount grown during capping was 60\,nm AlGaAs followed by 30\,nm GaAs. The total diameter of the nanowires is approximately 85\,nm. The structure has shown to be 100\% wurtzite with a few twin planes.\cite{ref25}

Single nanowire spectroscopy was realized on nanowires dispersed on a marked silicon substrate. In the Raman spectroscopy experiments, the nanowires were photoexcited by Ar$^+$Kr$^+$ or HeNe lasers with wavelengths respectively 647.1\,nm and 632.8\,nm. In the photoluminescence excitation spectroscopy measurements (PLE), the excitation source was a Koheras SuperK super continuum source filtered by an acusto-optical tunable filter (AOTF). During the PLE measurement the actual power of the excitation light was kept constant throughout the entire wavelength range by means of a computer controlled feedback loop. Both in the Raman and PLE spectroscopy experiments, the light was focused to a sub-micron spot using a cover glass corrected 0.75 NA microscope objective. The measurements were realized at a temperature between 10 and 360\,K in a liquid helium flow cryostat. The scattered light was collected through the same objective and focused on the entrance slit of a triple spectrometer and the spectrum collected thanks to a peltier cooled charge coupled device.

\begin{figure*} \includegraphics[width=\textwidth]{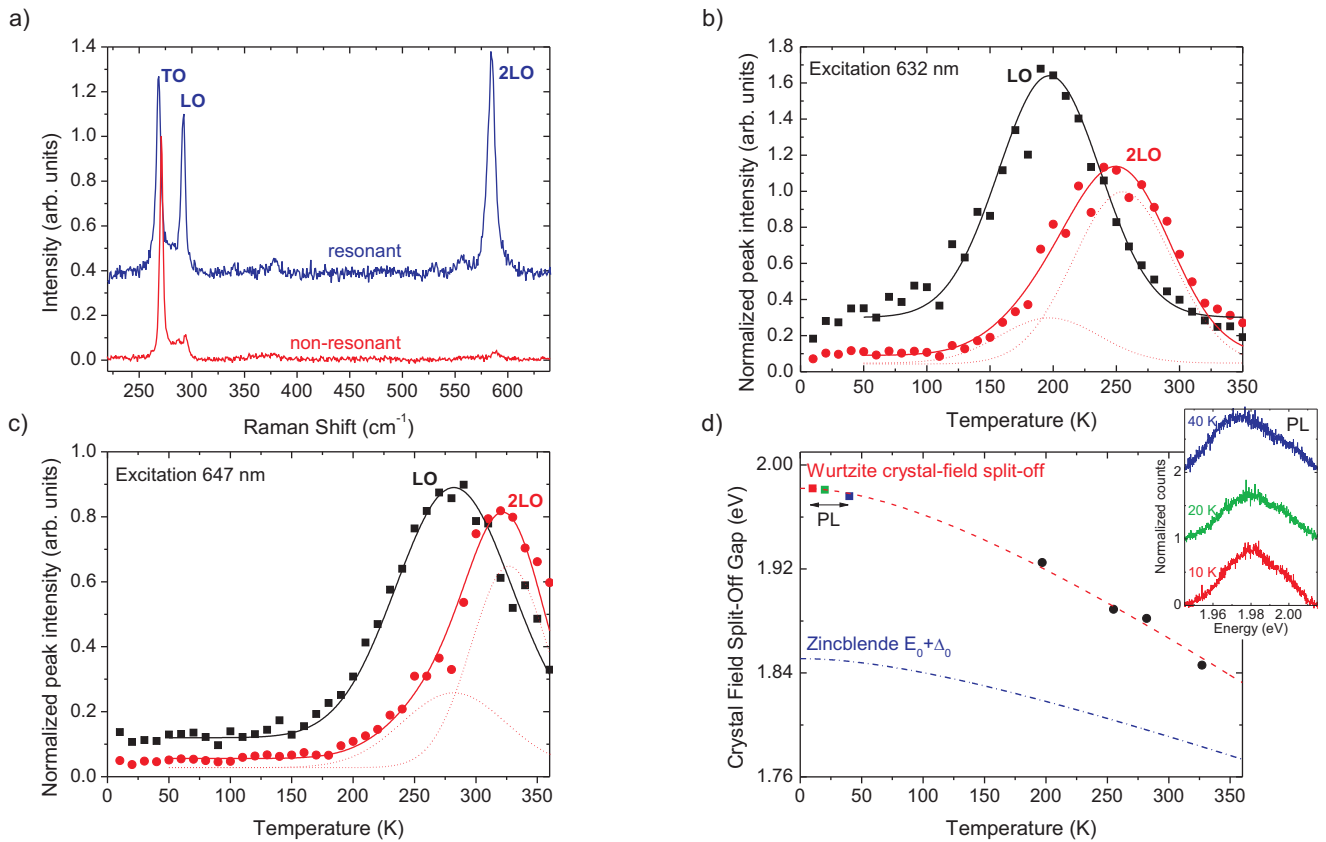}
\caption{\label{figure2} (a) Raman spectra of individual wurtzite GaAs nanowires at LO/2LO resonance and out of resonance. (b) and (c) show LO and 2LO resonance profiles for 632.8\,nm and 647.1\,nm excitation. The LO phonon scattering exhibits a single maximum under outgoing resonance. The 2LO resonance profile consists of a strong outgoing resonance ($E_{Laser}=E_c+\hbar \omega_{2LO}$) and a weaker intermediate resonance ($E_{Laser}=E_c+\hbar \omega_{LO}$). (d) Temperature dependent variation of the crystal field split-off gap in wurtzite GaAs with a fit to the Varshni equation. The inset shows the measured photoluminescence from this gap at three different temperatures. The temperature dependence of the zincblende $E_0+\Delta_0$ gap\cite{ref33} is shown for comparison.	
} \end{figure*}

We first present the resonant Raman scattering experiment. Here, we look for the conditions leading to sharp resonances of the first and second order LO phonons that occur via the dipole-forbidden Fr\"{o}hlich electron-phonon interaction.\cite{ref34,ref35} The resonance is observed when the excitation energy coincides with an interband critical point $E_c$ in the joint density of states of the semiconductor. In our case,  we reach the transition between the split-off valence and  the conduction band. Typical Raman spectra of wurtzite GaAs obtained in polarized configuration with the incident and detected polarization parallel to the c-axis, which lies along the nanowire axis, are shown in Fig.\,\ref{figure2}a. This configuration is denoted as $x(z,z)\bar{x}$ in Porto notation. We plot the spectra under non-resonant and resonant conditions, the difference being the intensity of the LO and 2LO peaks. Under non-resonant conditions, only the $A_1$(TO) mode at $\sim$270\,cm$^{-1}$ is allowed in $x(z,z)\bar{x}$ configuration.\cite{ref36,ref37} Under resonant conditions, not only the intensity of the dipole-forbidden $A_1$(LO) mode at $\sim$290\,cm$^{-1}$ increases significantly but also the second order Raman scattering by two $A_1$(LO)  phonons at $\sim$580\,cm$^{-1}$ is strongly enhanced.\cite{ref38} For simplicity, in the following we will denominate the $A_1$(TO) and $A_1$(LO) modes as simply TO and LO.

Now we proceed with the determination of the resonance Raman conditions for the measurement of the critical points of wurtzite GaAs. We measured the Raman spectra of single wurtzite GaAs nanowires as a function of the excitation energy and temperature. The excitation wavelengths used were 632.8 and 647.1\,nm. The temperature was varied between 10 and 360\,K. The intensity of the LO and 2LO peaks normalized to the intensity of the TO mode as a function of temperature for the excitation at 632.8 and 647.1\,nm are shown respectively in Fig.\,\ref{figure2}b and c. The resonance profile of the LO phonon scattering shows a single maximum under outgoing resonance, where the scattered light exactly matches a gap of the electronic band structure. The 2LO phonon scattering reveals a strong outgoing resonance ($E_{Laser}=E_c+\hbar \omega_{2LO}$) as well as a weaker intermediate resonance ($E_{Laser}=E_c+\hbar \omega_{LO}$). No incoming resonance is observed neither for the LO nor the 2LO scattering. For the excitation at 632.8\,nm, we observe the strongest resonance of the LO and 2LO peaks respectively at 197 and 255\,K. For the excitation at 647.1\,nm, we observe it at 282 and 327 K. For these temperatures, the energy of the critical point $E_c$ is then calculated:
\begin{equation}
E_c+ \hbar \omega_{Ph}=hc/\lambda
\end{equation}
where $\omega_{Ph}$ corresponds to the frequency of the phonons (LO or 2LO) and $\lambda$ is the excitation wavelength. For the temperatures of 197, 255, 282, and 327\,K, under which the resonances occur, we obtain respectively critical energies of 1.925, 1.889, 1.882, and 1.846\,eV. These points are reported in Fig.\,\ref{figure2}d. Limitations in wavelengths we have available do not allow us to obtain the energy of this transition at lower temperatures. Nevertheless, we have tried to measure direct luminescence from the recombination between the two resonant levels. Due to the very few non occupied states in the CH split-off band, such a transition is extremely weak. We have obtained luminescence of this transition for an incident polarization parallel to the hexagonal c-axis at temperatures between 10 and 40\,K by exciting with 568.2\,nm and a power of 50\,$\mu$W. The acquisition time was 30\,min, which is between three and four orders of magnitude longer than our typical luminescence experiments in our nanowires for equivalent excitation powers. The spectra are shown in the inset of Fig.\,\ref{figure2}d. At temperatures of 10, 20 and 40\,K we observe respectively PL centered at 1.982, 1.981, and 1.976\,eV. This enables us to complete the curve of the temperature dependence. The temperature dependent variation of the band gap energy can be commonly given in terms of the $\alpha$ and $\beta$ coefficients of the Varshni equation:\cite{ref39} 
\begin{equation}
E_c (T)=E_c (0)-\frac{\alpha T^2}{T+\beta}
\end{equation}

Least-squares fitting to the experimental data, the result is shown in Fig.\,\ref{figure2}d, gives the fitting parameters $\alpha$ and $\beta$ as $6.9 \cdot 10^{-4}$\,eV/K and 245.8\,K, respectively. For $T = 0$\,K we find a gap energy of $E_c(0) = 1.982$\,eV.
We now discuss the nature of the extrapolated interband critical point $E_c(T=0) = 1.982$\,eV in wurtzite GaAs. In zinc-blende GaAs, the interband transition from the spin-orbit-split valence band to the lowest conduction band at the $\Gamma$ point $E_0+\Delta_{0}$ is found at 1.851\,eV\cite{ref33} for $T = 0$\,K (see Fig.\,\ref{figure2}d). Likewise, we attribute the observed energy gap in wurtzite GaAs to a transition from the crystal-field split-off valence band to one of the lowest energy conduction bands at the $\Gamma$ point of the Brillouin zone. For the discussion, we need to come back to Fig.\,\ref{figure1}. The crystal-field split-off valence band is labeled with CH. This band has a $\Gamma_7$ symmetry. In the conduction band there is an important difference with respect to zinc-blende. There is one first minimum labeled $\Gamma_8$, which originates from the zone-folded L-valleys of zinc-blende GaAs. This band is separated by a small energy fraction $\Delta_{CB}$ from a close lying conduction band with $\Gamma_7$ symmetry. According to the selection rules in materials with hexagonal wurtzite structure, optical transitions from the $\Gamma_{7v}$ CH valence band to the $\Gamma_{7c}$ conduction band are dipole allowed. Transitions from the $\Gamma_{7v}$ valence band to the $\Gamma_{8c}$ conduction band are dipole forbidden.\cite{ref42} Generally, these selection rules may be softened in resonant Raman exciting conditions,\cite{ref40} meaning that resonant Raman scattering from an optically forbidden transition cannot be completely excluded. However, the fact that we observe photoluminescence from this energy gap let us conclude that the transition should be the dipole allowed CH ($\Gamma_{7v}$) to $\Gamma_{7c}$. Consequently, we assign the observed critical point with energy of 1.982\,eV in wurtzite GaAs to the interband transition from the crystal-field split-off valence band to the second lowest conduction band. Finally, we compare the experimental findings with theoretical predictions. Based on an empirical pseudopotential method including spin-orbit coupling, De und Pryor calculated values of respectively 1.978\,eV and 2.063eV for the $\Gamma_{7v}$ to $\Gamma_{8c}$ and $\Gamma_{7v}$ to $\Gamma_{7c}$ interband transitions.  This means that our experiment agrees within 4\% (81\,meV) with this theory.

\begin{figure} \includegraphics[width=\columnwidth]{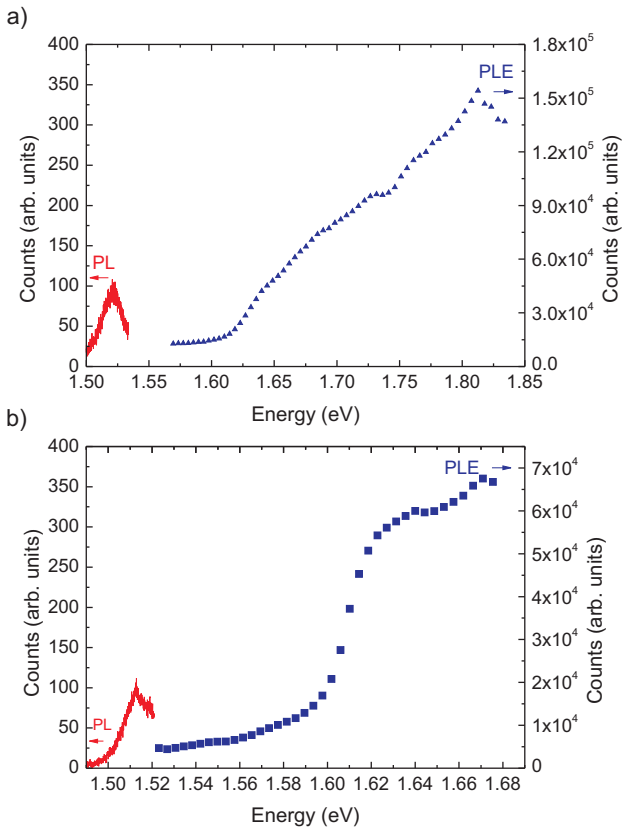}
\caption{\label{figure3}a) Integrated PLE signal data points (blue triangles) of a single nanowire. The red curve corresponds to the PL signal obtained for excitation with photon energy of 1.845\,eV b) Integrated PLE data from a nanowire ensemble (blue squares).  The red line shows the corresponding photoluminescence spectrum for excitation wavelength of 1.527\,eV
} \end{figure}
For a further understanding of the band structure of wurtzite GaAs, photoluminescence excitation spectroscopy was realized. A typical PL spectrum of a single nanowire is shown in Fig.\,\ref{figure3}a. A single peak centered at 1.515\,eV is observed, consistent with our previous works.\cite{ref25} The excitation spectroscopy measurements were realized by detecting the integrated intensity of the emission as a function the excitation energy. The resulting PLE measurements on single nanowires are shown in Fig.\,\ref{figure3}a. We start by describing the measurements realized at high photon energies of the excitation light, between 1.58 and 1.85\,eV, shown in Fig.\,\ref{figure3}a. The power density for these measurements was in the order of 300 W/cm$^2$. We observe a relatively sharp increase in the signal at $\sim$1.6\,eV. We believe the increased signal comes from a contribution of the {Al$_{0.33}$Ga$_{0.67}$As} shell, which exhibits about 2.7 times the volume of the nanowire core in the present sample. The electron-hole pairs generated in the shell can diffuse and recombine the wurtzite GaAs core, thereby contributing to the PLE signal. The bandgap of wurtzite-AlAs is theoretically expected to be significantly smaller compared to the (indirect band-gap) zinc-blende counterpart as a result of the zone-folding along the $\Gamma-L$ direction.\cite{ref19,ref20} A recombination around 1.6\,eV is consistent with recent measurements on wurtzite Al$_x$Ga$_{1-x}$As with comparable nominal composition.\cite{ref41} In principle this transition could also be in reasonable agreement with a transition from the heavy-hole band to the second conduction band with $\Gamma_7$ symmetry that is predicted at 1.588\,eV, and even with a transition from the $\Gamma_{7v}$ light-hole band to the $\Gamma_{8c}$ conduction band (1.623\,eV).\cite{ref19} However, according to theory, the transition $\Gamma_{7v}$ to $\Gamma_{8c}$ should not be dipole allowed\cite{ref42} and furthermore the transitions related to the light-hole band or the $\Gamma_7$ conduction band should be weak due to the smaller joint density of states.\cite{ref19} We therefore believe that the increase in PLE signal around 1.6\,eV signal is predominantly caused by the onset of absorption in the wurtzite {Al$_{0.33}$Ga$_{0.67}$As}  shell.

Finally, we turn our attention to PLE measurements realized closer to the bandgap. In order to approach the band gap with the excitation source the spectral output of the AOTF had to be further narrowed by passing the excitation light through a monochromator (f=300\,mm, Grating: 150\,Grooves/mm) and subsequently projecting the light from the exit slit of this monochromator to the sample. This enabled us to narrow down the linewidth below 1\,nm and to remove the remaining background emission from the AOTF. This further step also limited the maximum excitation power density down to about 10 W/cm$^2$. As a consequence, the collected signal was significantly reduced with respect to the measurements for excitation energies higher than 1.6\,eV. In order to increase the signal to noise ratio for the energy range between 1.525 and 1.68\,eV, we realized measurements on nanowire ensembles. As shown in Fig.\,\ref{figure3}b, the PL spectrum of the ensemble is extremely similar to that of single nanowires. This is possible because all nanowires present the identical structure. Consistent with the measurements at higher excitation energies, we observe an increase in PLE intensity around 1.6\,eV. For energies below 1.6\,eV and down to 1.523\,eV, no other clearly resolvable PLE feature is observed.  One should note that the PL spectrum shown in Fig.\,\ref{figure3}b corresponds to the one obtained at the excitation energy of 1.527\,eV. We therefore estimate the Stokes shift - the energy offset between the emission peak and the onset of absorption - to be smaller than 10\,meV. From our PLE measurements we can estimate an upper limit for the band gap of wurtzite GaAs to 1.523\,eV. As we have discussed in detail previously,\cite{ref25} taking into account quantum confinement effects, the emission peak is in reasonable agreement with predictions of De and Pryor of a bulk wurtzite band-gap of 1.503\,eV.\cite{ref19} At the same time, our PLE data are inconsistent with a band-gap of 1.552\,eV as predicted by Murayama and Nakayama,\cite{ref20} as no feature is observed in the corresponding spectral region.

In conclusion, we have measured the position of the crystal-field split-off band of wurtzite GaAs by resonant Raman and photoluminescence spectroscopy. The temperature dependence was fit with the Varshni equation and the parameters were extracted. A value of 1.982\,eV was obtained for the CH split-off to conduction band transition at 0 K. PLE measurements down to 1.525\,eV are consistent with a bandgap of wurtzite GaAs below 1.523\,eV and inconsistent with a transition at 1.55\,eV. 

\begin{acknowledgments}
 The authors thank A. Rudolph and D. Schuh for their experimental support. We thank financial support of by the Swiss National Science Foundation under Grant No. 2000021-121758/1 and 129775/1 by the European Research Council under Grant ‘Upcon’.
\end{acknowledgments}


\begin{thebibliography}{42}%
\makeatletter
\providecommand \@ifxundefined [1]{%
 \@ifx{#1\undefined}
}%
\providecommand \@ifnum [1]{%
 \ifnum #1\expandafter \@firstoftwo
 \else \expandafter \@secondoftwo
 \fi
}%
\providecommand \@ifx [1]{%
 \ifx #1\expandafter \@firstoftwo
 \else \expandafter \@secondoftwo
 \fi
}%
\providecommand \natexlab [1]{#1}%
\providecommand \enquote  [1]{``#1''}%
\providecommand \bibnamefont  [1]{#1}%
\providecommand \bibfnamefont [1]{#1}%
\providecommand \citenamefont [1]{#1}%
\providecommand \href@noop [0]{\@secondoftwo}%
\providecommand \href [0]{\begingroup \@sanitize@url \@href}%
\providecommand \@href[1]{\@@startlink{#1}\@@href}%
\providecommand \@@href[1]{\endgroup#1\@@endlink}%
\providecommand \@sanitize@url [0]{\catcode `\\12\catcode `\$12\catcode
  `\&12\catcode `\#12\catcode `\^12\catcode `\_12\catcode `\%12\relax}%
\providecommand \@@startlink[1]{}%
\providecommand \@@endlink[0]{}%
\providecommand \url  [0]{\begingroup\@sanitize@url \@url }%
\providecommand \@url [1]{\endgroup\@href {#1}{\urlprefix }}%
\providecommand \urlprefix  [0]{URL }%
\providecommand \Eprint [0]{\href }%
\providecommand \doibase [0]{http://dx.doi.org/}%
\providecommand \selectlanguage [0]{\@gobble}%
\providecommand \bibinfo  [0]{\@secondoftwo}%
\providecommand \bibfield  [0]{\@secondoftwo}%
\providecommand \translation [1]{[#1]}%
\providecommand \BibitemOpen [0]{}%
\providecommand \bibitemStop [0]{}%
\providecommand \bibitemNoStop [0]{.\EOS\space}%
\providecommand \EOS [0]{\spacefactor3000\relax}%
\providecommand \BibitemShut  [1]{\csname bibitem#1\endcsname}%
\let\auto@bib@innerbib\@empty
\bibitem [{\citenamefont {Qian}\ \emph {et~al.}(2008)\citenamefont {Qian},
  \citenamefont {Li}, \citenamefont {Grade\v{c}ak}, \citenamefont {Park},
  \citenamefont {Dong}, \citenamefont {Ding}, \citenamefont {Wang},\ and\
  \citenamefont {Lieber}}]{ref1}%
  \BibitemOpen
  \bibfield  {author} {\bibinfo {author} {\bibfnamefont {F.}~\bibnamefont
  {Qian}}, \bibinfo {author} {\bibfnamefont {Y.}~\bibnamefont {Li}}, \bibinfo
  {author} {\bibfnamefont {S.}~\bibnamefont {Grade\v{c}ak}}, \bibinfo {author}
  {\bibfnamefont {H.-G.}\ \bibnamefont {Park}}, \bibinfo {author}
  {\bibfnamefont {Y.}~\bibnamefont {Dong}}, \bibinfo {author} {\bibfnamefont
  {Y.}~\bibnamefont {Ding}}, \bibinfo {author} {\bibfnamefont {Z.~L.}\
  \bibnamefont {Wang}}, \ and\ \bibinfo {author} {\bibfnamefont {C.~M.}\
  \bibnamefont {Lieber}},\ }\href {\doibase 10.1038/nmat2253} {\bibfield
  {journal} {\bibinfo  {journal} {Nat Mater}\ }\textbf {\bibinfo {volume}
  {7}},\ \bibinfo {pages} {701} (\bibinfo {year} {2008})}\BibitemShut {NoStop}%
\bibitem [{\citenamefont {Qian}\ \emph {et~al.}(2004)\citenamefont {Qian},
  \citenamefont {Li}, \citenamefont {Grade\v{c}ak}, \citenamefont {Wang},
  \citenamefont {Barrelet},\ and\ \citenamefont {Lieber}}]{ref2}%
  \BibitemOpen
  \bibfield  {author} {\bibinfo {author} {\bibfnamefont {F.}~\bibnamefont
  {Qian}}, \bibinfo {author} {\bibfnamefont {Y.}~\bibnamefont {Li}}, \bibinfo
  {author} {\bibfnamefont {S.}~\bibnamefont {Grade\v{c}ak}}, \bibinfo {author}
  {\bibfnamefont {D.}~\bibnamefont {Wang}}, \bibinfo {author} {\bibfnamefont
  {C.~J.}\ \bibnamefont {Barrelet}}, \ and\ \bibinfo {author} {\bibfnamefont
  {C.~M.}\ \bibnamefont {Lieber}},\ }\href {\doibase 10.1021/nl0487774}
  {\bibfield  {journal} {\bibinfo  {journal} {Nano Letters}\ }\textbf {\bibinfo
  {volume} {4}},\ \bibinfo {pages} {1975} (\bibinfo {year} {2004})}\BibitemShut
  {NoStop}%
\bibitem [{\citenamefont {Xiang}\ \emph {et~al.}(2006)\citenamefont {Xiang},
  \citenamefont {Lu}, \citenamefont {Hu}, \citenamefont {Wu}, \citenamefont
  {Yan},\ and\ \citenamefont {Lieber}}]{ref3}%
  \BibitemOpen
  \bibfield  {author} {\bibinfo {author} {\bibfnamefont {J.}~\bibnamefont
  {Xiang}}, \bibinfo {author} {\bibfnamefont {W.}~\bibnamefont {Lu}}, \bibinfo
  {author} {\bibfnamefont {Y.}~\bibnamefont {Hu}}, \bibinfo {author}
  {\bibfnamefont {Y.}~\bibnamefont {Wu}}, \bibinfo {author} {\bibfnamefont
  {H.}~\bibnamefont {Yan}}, \ and\ \bibinfo {author} {\bibfnamefont {C.~M.}\
  \bibnamefont {Lieber}},\ }\href {\doibase 10.1038/nature04796} {\bibfield
  {journal} {\bibinfo  {journal} {Nature}\ }\textbf {\bibinfo {volume} {441}},\
  \bibinfo {pages} {489} (\bibinfo {year} {2006})}\BibitemShut {NoStop}%
\bibitem [{\citenamefont {Greytak}\ \emph {et~al.}(2005)\citenamefont
  {Greytak}, \citenamefont {Barrelet}, \citenamefont {Li},\ and\ \citenamefont
  {Lieber}}]{ref4}%
  \BibitemOpen
  \bibfield  {author} {\bibinfo {author} {\bibfnamefont {A.~B.}\ \bibnamefont
  {Greytak}}, \bibinfo {author} {\bibfnamefont {C.~J.}\ \bibnamefont
  {Barrelet}}, \bibinfo {author} {\bibfnamefont {Y.}~\bibnamefont {Li}}, \ and\
  \bibinfo {author} {\bibfnamefont {C.~M.}\ \bibnamefont {Lieber}},\ }\href
  {\doibase 10.1063/1.2089157} {\bibfield  {journal} {\bibinfo  {journal}
  {Applied Physics Letters}\ }\textbf {\bibinfo {volume} {87}},\ \bibinfo {eid}
  {151103} (\bibinfo {year} {2005})}\BibitemShut {NoStop}%
\bibitem [{\citenamefont {Fluegel}\ \emph {et~al.}(2007)\citenamefont
  {Fluegel}, \citenamefont {Mascarenhas}, \citenamefont {Snoke},\ and\
  \citenamefont {Pfeiffer}}]{ref5}%
  \BibitemOpen
  \bibfield  {author} {\bibinfo {author} {\bibfnamefont {B.}~\bibnamefont
  {Fluegel}}, \bibinfo {author} {\bibfnamefont {A.}~\bibnamefont
  {Mascarenhas}}, \bibinfo {author} {\bibfnamefont {D.~W.}\ \bibnamefont
  {Snoke}}, \ and\ \bibinfo {author} {\bibfnamefont {K.}~\bibnamefont
  {Pfeiffer}, \bibfnamefont {L.~N.and~West}},\ }\href {\doibase
  10.1038/nphoton.2007.229} {\bibfield  {journal} {\bibinfo  {journal} {Nat
  Photon}\ }\textbf {\bibinfo {volume} {1}},\ \bibinfo {pages} {701} (\bibinfo
  {year} {2007})}\BibitemShut {NoStop}%
\bibitem [{\citenamefont {Barnham}\ and\ \citenamefont {Duggan}(1990)}]{ref6}%
  \BibitemOpen
  \bibfield  {author} {\bibinfo {author} {\bibfnamefont {K.~W.~J.}\
  \bibnamefont {Barnham}}\ and\ \bibinfo {author} {\bibfnamefont
  {G.}~\bibnamefont {Duggan}},\ }\href {\doibase 10.1063/1.345339} {\bibfield
  {journal} {\bibinfo  {journal} {Journal of Applied Physics}\ }\textbf
  {\bibinfo {volume} {67}},\ \bibinfo {pages} {3490} (\bibinfo {year}
  {1990})}\BibitemShut {NoStop}%
\bibitem [{\citenamefont {Ferrari}\ \emph {et~al.}(2009)\citenamefont
  {Ferrari}, \citenamefont {Goldoni}, \citenamefont {Bertoni}, \citenamefont
  {Cuoghi},\ and\ \citenamefont {Molinari}}]{ref7}%
  \BibitemOpen
  \bibfield  {author} {\bibinfo {author} {\bibfnamefont {G.}~\bibnamefont
  {Ferrari}}, \bibinfo {author} {\bibfnamefont {G.}~\bibnamefont {Goldoni}},
  \bibinfo {author} {\bibfnamefont {A.}~\bibnamefont {Bertoni}}, \bibinfo
  {author} {\bibfnamefont {G.}~\bibnamefont {Cuoghi}}, \ and\ \bibinfo {author}
  {\bibfnamefont {E.}~\bibnamefont {Molinari}},\ }\href {\doibase
  10.1021/nl803942p} {\bibfield  {journal} {\bibinfo  {journal} {Nano Letters}\
  }\textbf {\bibinfo {volume} {9}},\ \bibinfo {pages} {1631} (\bibinfo {year}
  {2009})}\BibitemShut {NoStop}%
\bibitem [{\citenamefont {Glas}(2006)}]{ref8}%
  \BibitemOpen
  \bibfield  {author} {\bibinfo {author} {\bibfnamefont {F.}~\bibnamefont
  {Glas}},\ }\href {\doibase 10.1103/PhysRevB.74.121302} {\bibfield  {journal}
  {\bibinfo  {journal} {Phys. Rev. B}\ }\textbf {\bibinfo {volume} {74}},\
  \bibinfo {pages} {121302} (\bibinfo {year} {2006})}\BibitemShut {NoStop}%
\bibitem [{\citenamefont {M\r{a}rtensson}\ \emph {et~al.}(2004)\citenamefont
  {M\r{a}rtensson}, \citenamefont {Svensson}, \citenamefont {Wacaser},
  \citenamefont {Larsson}, \citenamefont {Seifert}, \citenamefont {Deppert},
  \citenamefont {Gustafsson}, \citenamefont {Wallenberg},\ and\ \citenamefont
  {Samuelson}}]{ref9}%
  \BibitemOpen
  \bibfield  {author} {\bibinfo {author} {\bibfnamefont {T.}~\bibnamefont
  {M\r{a}rtensson}}, \bibinfo {author} {\bibfnamefont {C.~P.~T.}\ \bibnamefont
  {Svensson}}, \bibinfo {author} {\bibfnamefont {B.~A.}\ \bibnamefont
  {Wacaser}}, \bibinfo {author} {\bibfnamefont {M.~W.}\ \bibnamefont
  {Larsson}}, \bibinfo {author} {\bibfnamefont {W.}~\bibnamefont {Seifert}},
  \bibinfo {author} {\bibfnamefont {K.}~\bibnamefont {Deppert}}, \bibinfo
  {author} {\bibfnamefont {A.}~\bibnamefont {Gustafsson}}, \bibinfo {author}
  {\bibfnamefont {L.~R.}\ \bibnamefont {Wallenberg}}, \ and\ \bibinfo {author}
  {\bibfnamefont {L.}~\bibnamefont {Samuelson}},\ }\href {\doibase
  10.1021/nl0487267} {\bibfield  {journal} {\bibinfo  {journal} {Nano Letters}\
  }\textbf {\bibinfo {volume} {4}},\ \bibinfo {pages} {1987} (\bibinfo {year}
  {2004})}\BibitemShut {NoStop}%
\bibitem [{\citenamefont {Wallentin}\ \emph {et~al.}(2010)\citenamefont
  {Wallentin}, \citenamefont {Persson}, \citenamefont {Wagner}, \citenamefont
  {Samuelson}, \citenamefont {Deppert},\ and\ \citenamefont
  {Borgstr\"{o}m}}]{ref10}%
  \BibitemOpen
  \bibfield  {author} {\bibinfo {author} {\bibfnamefont {J.}~\bibnamefont
  {Wallentin}}, \bibinfo {author} {\bibfnamefont {J.~M.}\ \bibnamefont
  {Persson}}, \bibinfo {author} {\bibfnamefont {J.~B.}\ \bibnamefont {Wagner}},
  \bibinfo {author} {\bibfnamefont {L.}~\bibnamefont {Samuelson}}, \bibinfo
  {author} {\bibfnamefont {K.}~\bibnamefont {Deppert}}, \ and\ \bibinfo
  {author} {\bibfnamefont {M.~T.}\ \bibnamefont {Borgstr\"{o}m}},\ }\href
  {\doibase 10.1021/nl903941b} {\bibfield  {journal} {\bibinfo  {journal} {Nano
  Letters}\ }\textbf {\bibinfo {volume} {10}},\ \bibinfo {pages} {974}
  (\bibinfo {year} {2010})}\BibitemShut {NoStop}%
\bibitem [{\citenamefont {Mattila}\ \emph {et~al.}(2006)\citenamefont
  {Mattila}, \citenamefont {Hakkarainen}, \citenamefont {Mulot},\ and\
  \citenamefont {Lipsanen}}]{ref11}%
  \BibitemOpen
  \bibfield  {author} {\bibinfo {author} {\bibfnamefont {M.}~\bibnamefont
  {Mattila}}, \bibinfo {author} {\bibfnamefont {T.}~\bibnamefont
  {Hakkarainen}}, \bibinfo {author} {\bibfnamefont {M.}~\bibnamefont {Mulot}},
  \ and\ \bibinfo {author} {\bibfnamefont {H.}~\bibnamefont {Lipsanen}},\
  }\href {\doibase 10.1088/0957-4484/17/6/008} {\bibfield  {journal} {\bibinfo
  {journal} {Nanotechnology}\ }\textbf {\bibinfo {volume} {17}},\ \bibinfo
  {pages} {1580} (\bibinfo {year} {2006})}\BibitemShut {NoStop}%
\bibitem [{\citenamefont {Mishra}\ \emph {et~al.}(2007)\citenamefont {Mishra},
  \citenamefont {Titova}, \citenamefont {Hoang}, \citenamefont {Jackson},
  \citenamefont {Smith}, \citenamefont {Yarrison-Rice}, \citenamefont {Kim},
  \citenamefont {Joyce}, \citenamefont {Gao}, \citenamefont {Tan},\ and\
  \citenamefont {Jagadish}}]{ref12}%
  \BibitemOpen
  \bibfield  {author} {\bibinfo {author} {\bibfnamefont {A.}~\bibnamefont
  {Mishra}}, \bibinfo {author} {\bibfnamefont {L.~V.}\ \bibnamefont {Titova}},
  \bibinfo {author} {\bibfnamefont {T.~B.}\ \bibnamefont {Hoang}}, \bibinfo
  {author} {\bibfnamefont {H.~E.}\ \bibnamefont {Jackson}}, \bibinfo {author}
  {\bibfnamefont {L.~M.}\ \bibnamefont {Smith}}, \bibinfo {author}
  {\bibfnamefont {J.~M.}\ \bibnamefont {Yarrison-Rice}}, \bibinfo {author}
  {\bibfnamefont {Y.}~\bibnamefont {Kim}}, \bibinfo {author} {\bibfnamefont
  {H.~J.}\ \bibnamefont {Joyce}}, \bibinfo {author} {\bibfnamefont
  {Q.}~\bibnamefont {Gao}}, \bibinfo {author} {\bibfnamefont {H.~H.}\
  \bibnamefont {Tan}}, \ and\ \bibinfo {author} {\bibfnamefont
  {C.}~\bibnamefont {Jagadish}},\ }\href {\doibase 10.1063/1.2828034}
  {\bibfield  {journal} {\bibinfo  {journal} {Applied Physics Letters}\
  }\textbf {\bibinfo {volume} {91}},\ \bibinfo {eid} {263104} (\bibinfo {year}
  {2007})}\BibitemShut {NoStop}%
\bibitem [{\citenamefont {Birman}(1959)}]{ref13}%
  \BibitemOpen
  \bibfield  {author} {\bibinfo {author} {\bibfnamefont {J.~L.}\ \bibnamefont
  {Birman}},\ }\href {\doibase 10.1103/PhysRevLett.2.157} {\bibfield  {journal}
  {\bibinfo  {journal} {Phys. Rev. Lett.}\ }\textbf {\bibinfo {volume} {2}},\
  \bibinfo {pages} {157} (\bibinfo {year} {1959})}\BibitemShut {NoStop}%
\bibitem [{\citenamefont {Algra}\ \emph {et~al.}(2008)\citenamefont {Algra},
  \citenamefont {Verheijen}, \citenamefont {Borgstrom}, \citenamefont {Feiner},
  \citenamefont {Immink}, \citenamefont {van Enckevort}, \citenamefont
  {Vlieg},\ and\ \citenamefont {Bakkers}}]{ref14}%
  \BibitemOpen
  \bibfield  {author} {\bibinfo {author} {\bibfnamefont {R.~E.}\ \bibnamefont
  {Algra}}, \bibinfo {author} {\bibfnamefont {M.~A.}\ \bibnamefont
  {Verheijen}}, \bibinfo {author} {\bibfnamefont {M.~T.}\ \bibnamefont
  {Borgstrom}}, \bibinfo {author} {\bibfnamefont {L.-F.}\ \bibnamefont
  {Feiner}}, \bibinfo {author} {\bibfnamefont {G.}~\bibnamefont {Immink}},
  \bibinfo {author} {\bibfnamefont {W.~J.~P.}\ \bibnamefont {van Enckevort}},
  \bibinfo {author} {\bibfnamefont {E.}~\bibnamefont {Vlieg}}, \ and\ \bibinfo
  {author} {\bibfnamefont {E.~P. A.~M.}\ \bibnamefont {Bakkers}},\ }\href
  {\doibase 10.1038/nature07570} {\bibfield  {journal} {\bibinfo  {journal}
  {Nature}\ }\textbf {\bibinfo {volume} {456}},\ \bibinfo {pages} {369}
  (\bibinfo {year} {2008})}\BibitemShut {NoStop}%
\bibitem [{\citenamefont {Caroff}\ \emph {et~al.}(2009)\citenamefont {Caroff},
  \citenamefont {Dick}, \citenamefont {Johansson}, \citenamefont {Messing},
  \citenamefont {Deppert},\ and\ \citenamefont {Samuelson}}]{ref15}%
  \BibitemOpen
  \bibfield  {author} {\bibinfo {author} {\bibfnamefont {P.}~\bibnamefont
  {Caroff}}, \bibinfo {author} {\bibfnamefont {K.~A.}\ \bibnamefont {Dick}},
  \bibinfo {author} {\bibfnamefont {J.}~\bibnamefont {Johansson}}, \bibinfo
  {author} {\bibfnamefont {M.~E.}\ \bibnamefont {Messing}}, \bibinfo {author}
  {\bibfnamefont {K.}~\bibnamefont {Deppert}}, \ and\ \bibinfo {author}
  {\bibfnamefont {L.}~\bibnamefont {Samuelson}},\ }\href {\doibase
  10.1038/nnano.2008.359} {\bibfield  {journal} {\bibinfo  {journal} {Nat
  Nano}\ }\textbf {\bibinfo {volume} {4}},\ \bibinfo {pages} {50} (\bibinfo
  {year} {2009})}\BibitemShut {NoStop}%
\bibitem [{\citenamefont {Dick}\ \emph {et~al.}(2010)\citenamefont {Dick},
  \citenamefont {Thelander}, \citenamefont {Samuelson},\ and\ \citenamefont
  {Caroff}}]{ref16}%
  \BibitemOpen
  \bibfield  {author} {\bibinfo {author} {\bibfnamefont {K.~A.}\ \bibnamefont
  {Dick}}, \bibinfo {author} {\bibfnamefont {C.}~\bibnamefont {Thelander}},
  \bibinfo {author} {\bibfnamefont {L.}~\bibnamefont {Samuelson}}, \ and\
  \bibinfo {author} {\bibfnamefont {P.}~\bibnamefont {Caroff}},\ }\href
  {\doibase 10.1021/nl101632a} {\bibfield  {journal} {\bibinfo  {journal} {Nano
  Letters}\ }\textbf {\bibinfo {volume} {10}},\ \bibinfo {pages} {3494}
  (\bibinfo {year} {2010})}\BibitemShut {NoStop}%
\bibitem [{\citenamefont {Perera}\ \emph {et~al.}(2010)\citenamefont {Perera},
  \citenamefont {Pemasiri}, \citenamefont {Fickenscher}, \citenamefont
  {Jackson}, \citenamefont {Smith}, \citenamefont {Yarrison-Rice},
  \citenamefont {Paiman}, \citenamefont {Gao}, \citenamefont {Tan},\ and\
  \citenamefont {Jagadish}}]{ref17}%
  \BibitemOpen
  \bibfield  {author} {\bibinfo {author} {\bibfnamefont {S.}~\bibnamefont
  {Perera}}, \bibinfo {author} {\bibfnamefont {K.}~\bibnamefont {Pemasiri}},
  \bibinfo {author} {\bibfnamefont {M.~A.}\ \bibnamefont {Fickenscher}},
  \bibinfo {author} {\bibfnamefont {H.~E.}\ \bibnamefont {Jackson}}, \bibinfo
  {author} {\bibfnamefont {L.~M.}\ \bibnamefont {Smith}}, \bibinfo {author}
  {\bibfnamefont {J.}~\bibnamefont {Yarrison-Rice}}, \bibinfo {author}
  {\bibfnamefont {S.}~\bibnamefont {Paiman}}, \bibinfo {author} {\bibfnamefont
  {Q.}~\bibnamefont {Gao}}, \bibinfo {author} {\bibfnamefont {H.~H.}\
  \bibnamefont {Tan}}, \ and\ \bibinfo {author} {\bibfnamefont
  {C.}~\bibnamefont {Jagadish}},\ }\href {\doibase 10.1063/1.3463036}
  {\bibfield  {journal} {\bibinfo  {journal} {Applied Physics Letters}\
  }\textbf {\bibinfo {volume} {97}},\ \bibinfo {eid} {023106} (\bibinfo {year}
  {2010})}\BibitemShut {NoStop}%
\bibitem [{\citenamefont {Gadret}\ \emph {et~al.}(2010)\citenamefont {Gadret},
  \citenamefont {Dias}, \citenamefont {Dacal}, \citenamefont {de~Lima},
  \citenamefont {Ruffo}, \citenamefont {Iikawa}, \citenamefont {Brasil},
  \citenamefont {Chiaramonte}, \citenamefont {Cotta}, \citenamefont {Tizei},
  \citenamefont {Ugarte},\ and\ \citenamefont {Cantarero}}]{ref18}%
  \BibitemOpen
  \bibfield  {author} {\bibinfo {author} {\bibfnamefont {E.~G.}\ \bibnamefont
  {Gadret}}, \bibinfo {author} {\bibfnamefont {G.~O.}\ \bibnamefont {Dias}},
  \bibinfo {author} {\bibfnamefont {L.~C.~O.}\ \bibnamefont {Dacal}}, \bibinfo
  {author} {\bibfnamefont {M.~M.}\ \bibnamefont {de~Lima}}, \bibinfo {author}
  {\bibfnamefont {C.~V. R.~S.}\ \bibnamefont {Ruffo}}, \bibinfo {author}
  {\bibfnamefont {F.}~\bibnamefont {Iikawa}}, \bibinfo {author} {\bibfnamefont
  {M.~J. S.~P.}\ \bibnamefont {Brasil}}, \bibinfo {author} {\bibfnamefont
  {T.}~\bibnamefont {Chiaramonte}}, \bibinfo {author} {\bibfnamefont {M.~A.}\
  \bibnamefont {Cotta}}, \bibinfo {author} {\bibfnamefont {L.~H.~G.}\
  \bibnamefont {Tizei}}, \bibinfo {author} {\bibfnamefont {D.}~\bibnamefont
  {Ugarte}}, \ and\ \bibinfo {author} {\bibfnamefont {A.}~\bibnamefont
  {Cantarero}},\ }\href {\doibase 10.1103/PhysRevB.82.125327} {\bibfield
  {journal} {\bibinfo  {journal} {Phys. Rev. B}\ }\textbf {\bibinfo {volume}
  {82}},\ \bibinfo {pages} {125327} (\bibinfo {year} {2010})}\BibitemShut
  {NoStop}%
\bibitem [{\citenamefont {De}\ and\ \citenamefont {Pryor}(2010)}]{ref19}%
  \BibitemOpen
  \bibfield  {author} {\bibinfo {author} {\bibfnamefont {A.}~\bibnamefont
  {De}}\ and\ \bibinfo {author} {\bibfnamefont {C.~E.}\ \bibnamefont {Pryor}},\
  }\href {\doibase 10.1103/PhysRevB.81.155210} {\bibfield  {journal} {\bibinfo
  {journal} {Phys. Rev. B}\ }\textbf {\bibinfo {volume} {81}},\ \bibinfo
  {pages} {155210} (\bibinfo {year} {2010})}\BibitemShut {NoStop}%
\bibitem [{\citenamefont {Murayama}\ and\ \citenamefont
  {Nakayama}(1994)}]{ref20}%
  \BibitemOpen
  \bibfield  {author} {\bibinfo {author} {\bibfnamefont {M.}~\bibnamefont
  {Murayama}}\ and\ \bibinfo {author} {\bibfnamefont {T.}~\bibnamefont
  {Nakayama}},\ }\href {\doibase 10.1103/PhysRevB.49.4710} {\bibfield
  {journal} {\bibinfo  {journal} {Phys. Rev. B}\ }\textbf {\bibinfo {volume}
  {49}},\ \bibinfo {pages} {4710} (\bibinfo {year} {1994})}\BibitemShut
  {NoStop}%
\bibitem [{\citenamefont {Hoang}\ \emph {et~al.}(2009)\citenamefont {Hoang},
  \citenamefont {Moses}, \citenamefont {Zhou}, \citenamefont {Dheeraj},
  \citenamefont {Fimland},\ and\ \citenamefont {Weman}}]{ref21}%
  \BibitemOpen
  \bibfield  {author} {\bibinfo {author} {\bibfnamefont {T.~B.}\ \bibnamefont
  {Hoang}}, \bibinfo {author} {\bibfnamefont {A.~F.}\ \bibnamefont {Moses}},
  \bibinfo {author} {\bibfnamefont {H.~L.}\ \bibnamefont {Zhou}}, \bibinfo
  {author} {\bibfnamefont {D.~L.}\ \bibnamefont {Dheeraj}}, \bibinfo {author}
  {\bibfnamefont {B.~O.}\ \bibnamefont {Fimland}}, \ and\ \bibinfo {author}
  {\bibfnamefont {H.}~\bibnamefont {Weman}},\ }\href {\doibase
  10.1063/1.3104853} {\bibfield  {journal} {\bibinfo  {journal} {Applied
  Physics Letters}\ }\textbf {\bibinfo {volume} {94}},\ \bibinfo {eid} {133105}
  (\bibinfo {year} {2009})}\BibitemShut {NoStop}%
\bibitem [{\citenamefont {Martelli}\ \emph {et~al.}(2007)\citenamefont
  {Martelli}, \citenamefont {Piccin}, \citenamefont {Bais}, \citenamefont
  {Jabeen}, \citenamefont {Ambrosini}, \citenamefont {Rubini},\ and\
  \citenamefont {Franciosi}}]{ref22}%
  \BibitemOpen
  \bibfield  {author} {\bibinfo {author} {\bibfnamefont {F.}~\bibnamefont
  {Martelli}}, \bibinfo {author} {\bibfnamefont {M.}~\bibnamefont {Piccin}},
  \bibinfo {author} {\bibfnamefont {G.}~\bibnamefont {Bais}}, \bibinfo {author}
  {\bibfnamefont {F.}~\bibnamefont {Jabeen}}, \bibinfo {author} {\bibfnamefont
  {S.}~\bibnamefont {Ambrosini}}, \bibinfo {author} {\bibfnamefont
  {S.}~\bibnamefont {Rubini}}, \ and\ \bibinfo {author} {\bibfnamefont
  {A.}~\bibnamefont {Franciosi}},\ }\href {\doibase
  10.1088/0957-4484/18/12/125603} {\bibfield  {journal} {\bibinfo  {journal}
  {Nanotechnology}\ }\textbf {\bibinfo {volume} {18}},\ \bibinfo {pages}
  {125603} (\bibinfo {year} {2007})}\BibitemShut {NoStop}%
\bibitem [{\citenamefont {Moewe}\ \emph {et~al.}(2008)\citenamefont {Moewe},
  \citenamefont {Chuang}, \citenamefont {Crankshaw}, \citenamefont {Chase},\
  and\ \citenamefont {Chang-Hasnain}}]{ref23}%
  \BibitemOpen
  \bibfield  {author} {\bibinfo {author} {\bibfnamefont {M.}~\bibnamefont
  {Moewe}}, \bibinfo {author} {\bibfnamefont {L.~C.}\ \bibnamefont {Chuang}},
  \bibinfo {author} {\bibfnamefont {S.}~\bibnamefont {Crankshaw}}, \bibinfo
  {author} {\bibfnamefont {C.}~\bibnamefont {Chase}}, \ and\ \bibinfo {author}
  {\bibfnamefont {C.}~\bibnamefont {Chang-Hasnain}},\ }\href {\doibase
  10.1063/1.2949315} {\bibfield  {journal} {\bibinfo  {journal} {Applied
  Physics Letters}\ }\textbf {\bibinfo {volume} {93}},\ \bibinfo {eid} {023116}
  (\bibinfo {year} {2008})}\BibitemShut {NoStop}%
\bibitem [{\citenamefont {Spirkoska}\ \emph {et~al.}(2009)\citenamefont
  {Spirkoska}, \citenamefont {Arbiol}, \citenamefont {Gustafsson},
  \citenamefont {Conesa-Boj}, \citenamefont {Glas}, \citenamefont {Zardo},
  \citenamefont {Heigoldt}, \citenamefont {Gass}, \citenamefont {Bleloch},
  \citenamefont {Estrade}, \citenamefont {Kaniber}, \citenamefont {Rossler},
  \citenamefont {Peiro}, \citenamefont {Morante}, \citenamefont {Abstreiter},
  \citenamefont {Samuelson},\ and\ \citenamefont {Fontcuberta~i
  Morral}}]{ref24}%
  \BibitemOpen
  \bibfield  {author} {\bibinfo {author} {\bibfnamefont {D.}~\bibnamefont
  {Spirkoska}}, \bibinfo {author} {\bibfnamefont {J.}~\bibnamefont {Arbiol}},
  \bibinfo {author} {\bibfnamefont {A.}~\bibnamefont {Gustafsson}}, \bibinfo
  {author} {\bibfnamefont {S.}~\bibnamefont {Conesa-Boj}}, \bibinfo {author}
  {\bibfnamefont {F.}~\bibnamefont {Glas}}, \bibinfo {author} {\bibfnamefont
  {I.}~\bibnamefont {Zardo}}, \bibinfo {author} {\bibfnamefont
  {M.}~\bibnamefont {Heigoldt}}, \bibinfo {author} {\bibfnamefont {M.~H.}\
  \bibnamefont {Gass}}, \bibinfo {author} {\bibfnamefont {A.~L.}\ \bibnamefont
  {Bleloch}}, \bibinfo {author} {\bibfnamefont {S.}~\bibnamefont {Estrade}},
  \bibinfo {author} {\bibfnamefont {M.}~\bibnamefont {Kaniber}}, \bibinfo
  {author} {\bibfnamefont {J.}~\bibnamefont {Rossler}}, \bibinfo {author}
  {\bibfnamefont {F.}~\bibnamefont {Peiro}}, \bibinfo {author} {\bibfnamefont
  {J.~R.}\ \bibnamefont {Morante}}, \bibinfo {author} {\bibfnamefont
  {G.}~\bibnamefont {Abstreiter}}, \bibinfo {author} {\bibfnamefont
  {L.}~\bibnamefont {Samuelson}}, \ and\ \bibinfo {author} {\bibfnamefont
  {A.}~\bibnamefont {Fontcuberta~i Morral}},\ }\href {\doibase
  10.1103/PhysRevB.80.245325} {\bibfield  {journal} {\bibinfo  {journal} {Phys.
  Rev. B}\ }\textbf {\bibinfo {volume} {80}},\ \bibinfo {pages} {245325}
  (\bibinfo {year} {2009})}\BibitemShut {NoStop}%
\bibitem [{\citenamefont {Heiss}\ \emph {et~al.}(2011)\citenamefont {Heiss},
  \citenamefont {Conesa-Boj}, \citenamefont {Ren}, \citenamefont {Tseng},
  \citenamefont {Gali}, \citenamefont {Rudolph}, \citenamefont {Uccelli},
  \citenamefont {Peir\'o}, \citenamefont {Morante}, \citenamefont {Schuh},
  \citenamefont {Reiger}, \citenamefont {Kaxiras}, \citenamefont {Arbiol},\
  and\ \citenamefont {Fontcuberta~i Morral}}]{ref25}%
  \BibitemOpen
  \bibfield  {author} {\bibinfo {author} {\bibfnamefont {M.}~\bibnamefont
  {Heiss}}, \bibinfo {author} {\bibfnamefont {S.}~\bibnamefont {Conesa-Boj}},
  \bibinfo {author} {\bibfnamefont {J.}~\bibnamefont {Ren}}, \bibinfo {author}
  {\bibfnamefont {H.-H.}\ \bibnamefont {Tseng}}, \bibinfo {author}
  {\bibfnamefont {A.}~\bibnamefont {Gali}}, \bibinfo {author} {\bibfnamefont
  {A.}~\bibnamefont {Rudolph}}, \bibinfo {author} {\bibfnamefont
  {E.}~\bibnamefont {Uccelli}}, \bibinfo {author} {\bibfnamefont
  {F.}~\bibnamefont {Peir\'o}}, \bibinfo {author} {\bibfnamefont {J.~R.}\
  \bibnamefont {Morante}}, \bibinfo {author} {\bibfnamefont {D.}~\bibnamefont
  {Schuh}}, \bibinfo {author} {\bibfnamefont {E.}~\bibnamefont {Reiger}},
  \bibinfo {author} {\bibfnamefont {E.}~\bibnamefont {Kaxiras}}, \bibinfo
  {author} {\bibfnamefont {J.}~\bibnamefont {Arbiol}}, \ and\ \bibinfo {author}
  {\bibfnamefont {A.}~\bibnamefont {Fontcuberta~i Morral}},\ }\href {\doibase
  10.1103/PhysRevB.83.045303} {\bibfield  {journal} {\bibinfo  {journal} {Phys.
  Rev. B}\ }\textbf {\bibinfo {volume} {83}},\ \bibinfo {pages} {045303}
  (\bibinfo {year} {2011})}\BibitemShut {NoStop}%
\bibitem [{\citenamefont {Chuang}\ and\ \citenamefont {Chang}(1996)}]{ref26}%
  \BibitemOpen
  \bibfield  {author} {\bibinfo {author} {\bibfnamefont {S.~L.}\ \bibnamefont
  {Chuang}}\ and\ \bibinfo {author} {\bibfnamefont {C.~S.}\ \bibnamefont
  {Chang}},\ }\href {\doibase 10.1103/PhysRevB.54.2491} {\bibfield  {journal}
  {\bibinfo  {journal} {Phys. Rev. B}\ }\textbf {\bibinfo {volume} {54}},\
  \bibinfo {pages} {2491} (\bibinfo {year} {1996})}\BibitemShut {NoStop}%
\bibitem [{\citenamefont {Pistol}\ \emph {et~al.}(1992)\citenamefont {Pistol},
  \citenamefont {Gerling}, \citenamefont {Hessman},\ and\ \citenamefont
  {Samuelson}}]{ref27}%
  \BibitemOpen
  \bibfield  {author} {\bibinfo {author} {\bibfnamefont {M.-E.}\ \bibnamefont
  {Pistol}}, \bibinfo {author} {\bibfnamefont {M.}~\bibnamefont {Gerling}},
  \bibinfo {author} {\bibfnamefont {D.}~\bibnamefont {Hessman}}, \ and\
  \bibinfo {author} {\bibfnamefont {L.}~\bibnamefont {Samuelson}},\ }\href
  {\doibase 10.1103/PhysRevB.45.3628} {\bibfield  {journal} {\bibinfo
  {journal} {Phys. Rev. B}\ }\textbf {\bibinfo {volume} {45}},\ \bibinfo
  {pages} {3628} (\bibinfo {year} {1992})}\BibitemShut {NoStop}%
\bibitem [{\citenamefont {Sk\"{o}ld}\ \emph {et~al.}(2009)\citenamefont
  {Sk\"{o}ld}, \citenamefont {Pistol}, \citenamefont {Dick}, \citenamefont
  {Pryor}, \citenamefont {Wagner}, \citenamefont {Karlsson},\ and\
  \citenamefont {Samuelson}}]{ref28}%
  \BibitemOpen
  \bibfield  {author} {\bibinfo {author} {\bibfnamefont {N.}~\bibnamefont
  {Sk\"{o}ld}}, \bibinfo {author} {\bibfnamefont {M.-E.}\ \bibnamefont
  {Pistol}}, \bibinfo {author} {\bibfnamefont {K.~A.}\ \bibnamefont {Dick}},
  \bibinfo {author} {\bibfnamefont {C.}~\bibnamefont {Pryor}}, \bibinfo
  {author} {\bibfnamefont {J.~B.}\ \bibnamefont {Wagner}}, \bibinfo {author}
  {\bibfnamefont {L.~S.}\ \bibnamefont {Karlsson}}, \ and\ \bibinfo {author}
  {\bibfnamefont {L.}~\bibnamefont {Samuelson}},\ }\href {\doibase
  10.1103/PhysRevB.80.041312} {\bibfield  {journal} {\bibinfo  {journal} {Phys.
  Rev. B}\ }\textbf {\bibinfo {volume} {80}},\ \bibinfo {pages} {041312}
  (\bibinfo {year} {2009})}\BibitemShut {NoStop}%
\bibitem [{\citenamefont {Calleja}\ and\ \citenamefont
  {Cardona}(1977)}]{ref29}%
  \BibitemOpen
  \bibfield  {author} {\bibinfo {author} {\bibfnamefont {J.~M.}\ \bibnamefont
  {Calleja}}\ and\ \bibinfo {author} {\bibfnamefont {M.}~\bibnamefont
  {Cardona}},\ }\href {\doibase 10.1103/PhysRevB.16.3753} {\bibfield  {journal}
  {\bibinfo  {journal} {Phys. Rev. B}\ }\textbf {\bibinfo {volume} {16}},\
  \bibinfo {pages} {3753} (\bibinfo {year} {1977})}\BibitemShut {NoStop}%
\bibitem [{\citenamefont {Wagner}\ and\ \citenamefont {Ellis}(1964)}]{ref30}%
  \BibitemOpen
  \bibfield  {author} {\bibinfo {author} {\bibfnamefont {R.~S.}\ \bibnamefont
  {Wagner}}\ and\ \bibinfo {author} {\bibfnamefont {W.~C.}\ \bibnamefont
  {Ellis}},\ }\href {\doibase 10.1063/1.1753975} {\bibfield  {journal}
  {\bibinfo  {journal} {Applied Physics Letters}\ }\textbf {\bibinfo {volume}
  {4}},\ \bibinfo {pages} {89} (\bibinfo {year} {1964})}\BibitemShut {NoStop}%
\bibitem [{\citenamefont {Rudolph}\ \emph {et~al.}(2009)\citenamefont
  {Rudolph}, \citenamefont {Soda}, \citenamefont {Kiessling}, \citenamefont
  {Wojtowicz}, \citenamefont {Schuh}, \citenamefont {Wegscheider},
  \citenamefont {Zweck}, \citenamefont {Back},\ and\ \citenamefont
  {Reiger}}]{ref31}%
  \BibitemOpen
  \bibfield  {author} {\bibinfo {author} {\bibfnamefont {A.}~\bibnamefont
  {Rudolph}}, \bibinfo {author} {\bibfnamefont {M.}~\bibnamefont {Soda}},
  \bibinfo {author} {\bibfnamefont {M.}~\bibnamefont {Kiessling}}, \bibinfo
  {author} {\bibfnamefont {T.}~\bibnamefont {Wojtowicz}}, \bibinfo {author}
  {\bibfnamefont {D.}~\bibnamefont {Schuh}}, \bibinfo {author} {\bibfnamefont
  {W.}~\bibnamefont {Wegscheider}}, \bibinfo {author} {\bibfnamefont
  {J.}~\bibnamefont {Zweck}}, \bibinfo {author} {\bibfnamefont
  {C.}~\bibnamefont {Back}}, \ and\ \bibinfo {author} {\bibfnamefont
  {E.}~\bibnamefont {Reiger}},\ }\href {\doibase 10.1021/nl9020717} {\bibfield
  {journal} {\bibinfo  {journal} {Nano Letters}\ }\textbf {\bibinfo {volume}
  {9}},\ \bibinfo {pages} {3860} (\bibinfo {year} {2009})}\BibitemShut
  {NoStop}%
\bibitem [{\citenamefont {Fontcuberta~i Morral}\ \emph
  {et~al.}(2008)\citenamefont {Fontcuberta~i Morral}, \citenamefont
  {Spirkoska}, \citenamefont {Arbiol}, \citenamefont {Heigoldt}, \citenamefont
  {Morante},\ and\ \citenamefont {Abstreiter}}]{ref32}%
  \BibitemOpen
  \bibfield  {author} {\bibinfo {author} {\bibfnamefont {A.}~\bibnamefont
  {Fontcuberta~i Morral}}, \bibinfo {author} {\bibfnamefont {D.}~\bibnamefont
  {Spirkoska}}, \bibinfo {author} {\bibfnamefont {J.}~\bibnamefont {Arbiol}},
  \bibinfo {author} {\bibfnamefont {M.}~\bibnamefont {Heigoldt}}, \bibinfo
  {author} {\bibfnamefont {J.~R.}\ \bibnamefont {Morante}}, \ and\ \bibinfo
  {author} {\bibfnamefont {G.}~\bibnamefont {Abstreiter}},\ }\href {\doibase
  10.1002/smll.200701091} {\bibfield  {journal} {\bibinfo  {journal} {Small}\
  }\textbf {\bibinfo {volume} {4}},\ \bibinfo {pages} {899} (\bibinfo {year}
  {2008})}\BibitemShut {NoStop}%
\bibitem [{\citenamefont {Lautenschlager}\ \emph {et~al.}(1987)\citenamefont
  {Lautenschlager}, \citenamefont {Garriga}, \citenamefont {Logothetidis},\
  and\ \citenamefont {Cardona}}]{ref33}%
  \BibitemOpen
  \bibfield  {author} {\bibinfo {author} {\bibfnamefont {P.}~\bibnamefont
  {Lautenschlager}}, \bibinfo {author} {\bibfnamefont {M.}~\bibnamefont
  {Garriga}}, \bibinfo {author} {\bibfnamefont {S.}~\bibnamefont
  {Logothetidis}}, \ and\ \bibinfo {author} {\bibfnamefont {M.}~\bibnamefont
  {Cardona}},\ }\href {\doibase 10.1103/PhysRevB.35.9174} {\bibfield  {journal}
  {\bibinfo  {journal} {Phys. Rev. B}\ }\textbf {\bibinfo {volume} {35}},\
  \bibinfo {pages} {9174} (\bibinfo {year} {1987})}\BibitemShut {NoStop}%
\bibitem [{\citenamefont {Trommer}\ and\ \citenamefont
  {Cardona}(1978)}]{ref34}%
  \BibitemOpen
  \bibfield  {author} {\bibinfo {author} {\bibfnamefont {R.}~\bibnamefont
  {Trommer}}\ and\ \bibinfo {author} {\bibfnamefont {M.}~\bibnamefont
  {Cardona}},\ }\href {\doibase 10.1103/PhysRevB.17.1865} {\bibfield  {journal}
  {\bibinfo  {journal} {Phys. Rev. B}\ }\textbf {\bibinfo {volume} {17}},\
  \bibinfo {pages} {1865} (\bibinfo {year} {1978})}\BibitemShut {NoStop}%
\bibitem [{\citenamefont {Kauschke}\ \emph {et~al.}(1987)\citenamefont
  {Kauschke}, \citenamefont {Cardona},\ and\ \citenamefont {Bauser}}]{ref35}%
  \BibitemOpen
  \bibfield  {author} {\bibinfo {author} {\bibfnamefont {W.}~\bibnamefont
  {Kauschke}}, \bibinfo {author} {\bibfnamefont {M.}~\bibnamefont {Cardona}}, \
  and\ \bibinfo {author} {\bibfnamefont {E.}~\bibnamefont {Bauser}},\ }\href
  {\doibase 10.1103/PhysRevB.35.8030} {\bibfield  {journal} {\bibinfo
  {journal} {Phys. Rev. B}\ }\textbf {\bibinfo {volume} {35}},\ \bibinfo
  {pages} {8030} (\bibinfo {year} {1987})}\BibitemShut {NoStop}%
\bibitem [{\citenamefont {Zardo}\ \emph {et~al.}(2009)\citenamefont {Zardo},
  \citenamefont {Conesa-Boj}, \citenamefont {Peiro}, \citenamefont {Morante},
  \citenamefont {Arbiol}, \citenamefont {Uccelli}, \citenamefont {Abstreiter},\
  and\ \citenamefont {Fontcuberta~i Morral}}]{ref36}%
  \BibitemOpen
  \bibfield  {author} {\bibinfo {author} {\bibfnamefont {I.}~\bibnamefont
  {Zardo}}, \bibinfo {author} {\bibfnamefont {S.}~\bibnamefont {Conesa-Boj}},
  \bibinfo {author} {\bibfnamefont {F.}~\bibnamefont {Peiro}}, \bibinfo
  {author} {\bibfnamefont {J.~R.}\ \bibnamefont {Morante}}, \bibinfo {author}
  {\bibfnamefont {J.}~\bibnamefont {Arbiol}}, \bibinfo {author} {\bibfnamefont
  {E.}~\bibnamefont {Uccelli}}, \bibinfo {author} {\bibfnamefont
  {G.}~\bibnamefont {Abstreiter}}, \ and\ \bibinfo {author} {\bibfnamefont
  {A.}~\bibnamefont {Fontcuberta~i Morral}},\ }\href {\doibase
  10.1103/PhysRevB.80.245324} {\bibfield  {journal} {\bibinfo  {journal} {Phys.
  Rev. B}\ }\textbf {\bibinfo {volume} {80}},\ \bibinfo {pages} {245324}
  (\bibinfo {year} {2009})}\BibitemShut {NoStop}%
\bibitem [{\citenamefont {Crankshaw}\ \emph {et~al.}(2010)\citenamefont
  {Crankshaw}, \citenamefont {Chuang}, \citenamefont {Moewe},\ and\
  \citenamefont {Chang-Hasnain}}]{ref37}%
  \BibitemOpen
  \bibfield  {author} {\bibinfo {author} {\bibfnamefont {S.}~\bibnamefont
  {Crankshaw}}, \bibinfo {author} {\bibfnamefont {L.~C.}\ \bibnamefont
  {Chuang}}, \bibinfo {author} {\bibfnamefont {M.}~\bibnamefont {Moewe}}, \
  and\ \bibinfo {author} {\bibfnamefont {C.}~\bibnamefont {Chang-Hasnain}},\
  }\href {\doibase 10.1103/PhysRevB.81.233303} {\bibfield  {journal} {\bibinfo
  {journal} {Phys. Rev. B}\ }\textbf {\bibinfo {volume} {81}},\ \bibinfo
  {pages} {233303} (\bibinfo {year} {2010})}\BibitemShut {NoStop}%
\bibitem [{\citenamefont {Brewster}\ \emph {et~al.}(2009)\citenamefont
  {Brewster}, \citenamefont {Schimek}, \citenamefont {Reich},\ and\
  \citenamefont {Grade\ifmmode~\check{c}\else \v{c}\fi{}ak}}]{ref38}%
  \BibitemOpen
  \bibfield  {author} {\bibinfo {author} {\bibfnamefont {M.}~\bibnamefont
  {Brewster}}, \bibinfo {author} {\bibfnamefont {O.}~\bibnamefont {Schimek}},
  \bibinfo {author} {\bibfnamefont {S.}~\bibnamefont {Reich}}, \ and\ \bibinfo
  {author} {\bibfnamefont {S.}~\bibnamefont {Grade\ifmmode~\check{c}\else
  \v{c}\fi{}ak}},\ }\href {\doibase 10.1103/PhysRevB.80.201314} {\bibfield
  {journal} {\bibinfo  {journal} {Phys. Rev. B}\ }\textbf {\bibinfo {volume}
  {80}},\ \bibinfo {pages} {201314} (\bibinfo {year} {2009})}\BibitemShut
  {NoStop}%
\bibitem [{\citenamefont {Varshni}(1967)}]{ref39}%
  \BibitemOpen
  \bibfield  {author} {\bibinfo {author} {\bibfnamefont {Y.~P.}\ \bibnamefont
  {Varshni}},\ }\href {\doibase 10.1016/0031-8914(67)90062-6} {\bibfield
  {journal} {\bibinfo  {journal} {Physica}\ }\textbf {\bibinfo {volume} {34}},\
  \bibinfo {pages} {149 } (\bibinfo {year} {1967})}\BibitemShut {NoStop}%
\bibitem [{\citenamefont {Tronc}\ \emph {et~al.}(1999)\citenamefont {Tronc},
  \citenamefont {Kitaev}, \citenamefont {Wang}, \citenamefont {Limonov},
  \citenamefont {Panfilov},\ and\ \citenamefont {Neu}}]{ref42}%
  \BibitemOpen
  \bibfield  {author} {\bibinfo {author} {\bibfnamefont {P.}~\bibnamefont
  {Tronc}}, \bibinfo {author} {\bibfnamefont {Y.}~\bibnamefont {Kitaev}},
  \bibinfo {author} {\bibfnamefont {G.}~\bibnamefont {Wang}}, \bibinfo {author}
  {\bibfnamefont {M.}~\bibnamefont {Limonov}}, \bibinfo {author} {\bibfnamefont
  {A.}~\bibnamefont {Panfilov}}, \ and\ \bibinfo {author} {\bibfnamefont
  {G.}~\bibnamefont {Neu}},\ }\href {\doibase
  10.1002/(SICI)1521-3951(199911)216:1<599::AID-PSSB599>3.0.CO;2-H} {\bibfield
  {journal} {\bibinfo  {journal} {phys. stat. sol. (b)}\ }\textbf {\bibinfo
  {volume} {216}},\ \bibinfo {pages} {599} (\bibinfo {year}
  {1999})}\BibitemShut {NoStop}%
\bibitem [{\citenamefont {Martin}\ and\ \citenamefont {Falicov}(1983)}]{ref40}%
  \BibitemOpen
  \bibfield  {author} {\bibinfo {author} {\bibfnamefont {R.}~\bibnamefont
  {Martin}}\ and\ \bibinfo {author} {\bibfnamefont {L.}~\bibnamefont
  {Falicov}},\ }in\ \href {\doibase 10.1007/3-540-11913-2_3} {\emph {\bibinfo
  {booktitle} {Light Scattering in Solids I}}},\ \bibinfo {series} {Topics in
  Applied Physics}, Vol.~\bibinfo {volume} {8}\ (\bibinfo  {publisher}
  {Springer Berlin / Heidelberg},\ \bibinfo {year} {1983})\ pp.\ \bibinfo
  {pages} {79--145}\BibitemShut {NoStop}%
\bibitem [{\citenamefont {Zhou}\ \emph {et~al.}(2009)\citenamefont {Zhou},
  \citenamefont {Hoang}, \citenamefont {Dheeraj}, \citenamefont {van Helvoort},
  \citenamefont {Liu}, \citenamefont {Harmand}, \citenamefont {Fimland},\ and\
  \citenamefont {Weman}}]{ref41}%
  \BibitemOpen
  \bibfield  {author} {\bibinfo {author} {\bibfnamefont {H.~L.}\ \bibnamefont
  {Zhou}}, \bibinfo {author} {\bibfnamefont {T.~B.}\ \bibnamefont {Hoang}},
  \bibinfo {author} {\bibfnamefont {D.~L.}\ \bibnamefont {Dheeraj}}, \bibinfo
  {author} {\bibfnamefont {A.~T.~J.}\ \bibnamefont {van Helvoort}}, \bibinfo
  {author} {\bibfnamefont {L.}~\bibnamefont {Liu}}, \bibinfo {author}
  {\bibfnamefont {J.~C.}\ \bibnamefont {Harmand}}, \bibinfo {author}
  {\bibfnamefont {B.~O.}\ \bibnamefont {Fimland}}, \ and\ \bibinfo {author}
  {\bibfnamefont {H.}~\bibnamefont {Weman}},\ }\href {\doibase
  10.1088/0957-4484/20/41/415701} {\bibfield  {journal} {\bibinfo  {journal}
  {Nanotechnology}\ }\textbf {\bibinfo {volume} {20}},\ \bibinfo {pages}
  {415701} (\bibinfo {year} {2009})}\BibitemShut {NoStop}%
\end{thebibliography}
\end{document}